\documentclass{jfm}
\usepackage{amssymb} 
\usepackage{mathtools}
\usepackage{epstopdf}
\usepackage[colon,square,sectionbib]{natbib}
\usepackage[colorlinks=true]{hyperref}
\hypersetup{
    colorlinks = true,
    citecolor=blue
}
\usepackage{pdfpages}
\usepackage{wrapfig}
\usepackage{enumitem}

\newcommand{\ce}{{\rm ce}}
\newcommand{\se}{{\rm se}}

\begin{document}

\shorttitle{Shear Dispersion} 
\shortauthor{M. A. Jimenez-Urias and T. W. N. Haine} 


\title{An exact solution to dispersion of a passive scalar by a periodic shear flow}

\author
 {
 Miguel A. Jimenez-Urias 
  \corresp{\email{mjimen17@jh.edu}}
\and
Thomas W. N. Haine
  }

\affiliation
{
Earth and Planetary Sciences, Johns Hopkins University, Baltimore, MD 21218, USA
}

\maketitle

\begin{abstract}

We present an exact analytical solution to the problem of shear dispersion given a general initial condition. The solution is expressed as an infinite series expansion involving Mathieu functions and their eigenvalues. The eigenvalue system depends on the imaginary parameter $q=2ik\Pen$, with $k$ the wavenumber that determines the tracer scale in the initial condition and $\Pen$ the P\'{e}clet number. Solutions are valid for all $\Pen$, $t>0$, and $k>0$ except at specific values of $q=q_{\ell}^{EP}$ called Exceptional Points (EPs), the first occurring at $q_{0}^{EP}\approx1.468i$. For values of $q \lessapprox 1.468i$, all the eigenvalues are real, different and eigenfunctions decay with time, thus shear dispersion can be represented as a diffusive process. For values of $q \gtrapprox 1.468i$, pairs of eigenvalues coalesce at EPs becoming complex conjugates, the eigenfunctions propagate and decay with time, and so shear dispersion is no longer a purely diffusive process.

The limit $q\rightarrow0$ is approached by the small P\'{e}clet number limit for all finite $k>0$, or equally by the large P\'{e}clet number limit as long as $2k \ll 1/\Pen$. The latter implies $k\rightarrow0$, strong separation of scales between the tracer and flow. The limit $q\rightarrow\infty$ results from large P\'{e}clet number for any $k>0$, or from large $k$ and non-vanishing $\Pen$.

We derive an exact closure that is continuous in wavenumber space. At small $q$, the closure approaches a diffusion operator with an effective diffusivity proportional to $U_0^2/\kappa$, for flow speed $U_0$ and diffusivity $\kappa$. At large $q$, the closure approaches the sum of an advection operator plus a half-derivative operator (differential operator of fractional order), the latter with coefficient proportional to $\sqrt{\kappa U_0}$.
\end{abstract}

\section{Introduction}

\textit{Shear dispersion} describes the enhanced spreading of a conservative passive scalar tracer due to the interaction of velocity gradients with weak diffusivities \citep{taylor1953dispersion, batchelor1956steady, aris1956dispersion, young1991shear}. Approximate models of shear dispersion suggest that the problem can be split into two stages \citep{rhines1983rapidly}. First, there is a \textit{rapid stage} during which stirring by the spatially-varying flow replaces the initial condition by a \textit{generalized average} along a streamline. Second, there is a \textit{slower stage} during which diffusive mixing takes place with an \textit{effective diffusivity} $\kappa^*\propto U_0^2/\kappa$, where $U_0$ is the characteristic flow speed and $\kappa$ is the molecular diffusivity. The enhancement of diffusivity by shear dispersion, which is strongest for small $\kappa$, is referred to as G. I. Taylor's {\it effective diffusivity} $\kappa^*$ \citep{taylor1953dispersion}.

Shear dispersion is often regarded as a special case of the method of homogenization (\citealt{haynes2014dispersion}; \citealt{hinchperturbation} Ch. 7; \citealt{holmes2012introduction}, Ch. 5). Homogenization of periodic flows exploits the separation of scales between the (small) scale of the flow and the (large) scale of the tracer, thus it is restricted to timescales longer than the diffusive timescale and so it fails to accurately represent the early-time behavior. Despite the lack of a solution for all times $t>0$, the effective diffusivity $\kappa^*$ is generally regarded as valid in the large P\'{e}clet number limit (although claims have been made that it applies only for small P\'{e}clet number; see \citealt{young1991shear} for a brief discussion). In the large P\'{e}clet number limit, approximate asymptotic solutions have been derived to extend the effective diffusivity, which involve higher order corrections that attempt to model early-time behavior \citep{mercer1990centre, young1991shear, haynes2014dispersion}. All these approximate solutions rely on homogenization theory, however, and thus fail to describe the complete spatial evolution of the tracer for all $t>0$.

Development of closures for the advection-diffusion equation remains an active area of research, particularly in climate modeling with the need to parameterize the effects of subgrid physics on the larger scale fields \citep{gent1990isopycnal, gent1995parameterizing}. An example is the parameterization of (subgrid) stirring by oceanic eddies on the distribution of salinity, temperature and nutrients in climate simulations. In analogy with shear dispersion, a widely used approach is to assume scale separation between the eddies and the slowly-evolving large-scale flow \citep{gent1990isopycnal, gent1995parameterizing, grooms2012multiscale, marshall2012closure}. Moreover, numerical mixing is often treated as isotropic down-gradient diffusion (along isentropes), even though it only applies to isotropic flows with Gaussian perturbations \citep{smith2005tracer}. A novel approach in turbulence modeling represents the subgrid turbulence processes using fractional order differential operators \citep{chen2006speculative, lischke2020fractional}. Such operators lie between differential and integral operators and are thus inherently non-local \citep{samko1993fractional}. There is recent indication that these operators can capture aspects of both the long and short time evolution of shear dispersion \citep{park2018macroscopic, mani2019macroscopic}, and thus can potentially provide a pathway for developing non-local and scale-dependent closures that expand beyond the limits of homogenization theory.

In this paper, we begin in section \ref{sec:Shear_Dispersion} by deriving an exact analytical solution to the problem of scalar dispersion by a periodic shear flow. The solution is valid across all times $t>0$, $\Pen$ and wavenumbers $k$ (except at combinations of $k$ and $\Pen$ that result in specific values of their product). The solution is obtained in accordance to Floquet theory, and it depends on a single parameter $q$, which is a function of the P\'{e}clet number and the wavenumbers of the initial condition. With the exact solution, we derive in section \ref{sec:Closures} the exact closure to the advective term that is continuous across wavenumbers and we derive approximations to the exact closure based on its dependence on $q$. These closures clarify the limits of validity of Taylor's effective diffusivity and reveal a novel regime. A discussion of implications and future opportunities is in section \ref{sec:discussion} and conclusions are in section \ref{sec:conclusion}.

\section{Plane Shear Dispersion}\label{sec:Shear_Dispersion}
\subsection{Problem Statement}
\begin{figure}
    \centering
    \includegraphics[width=250pt]{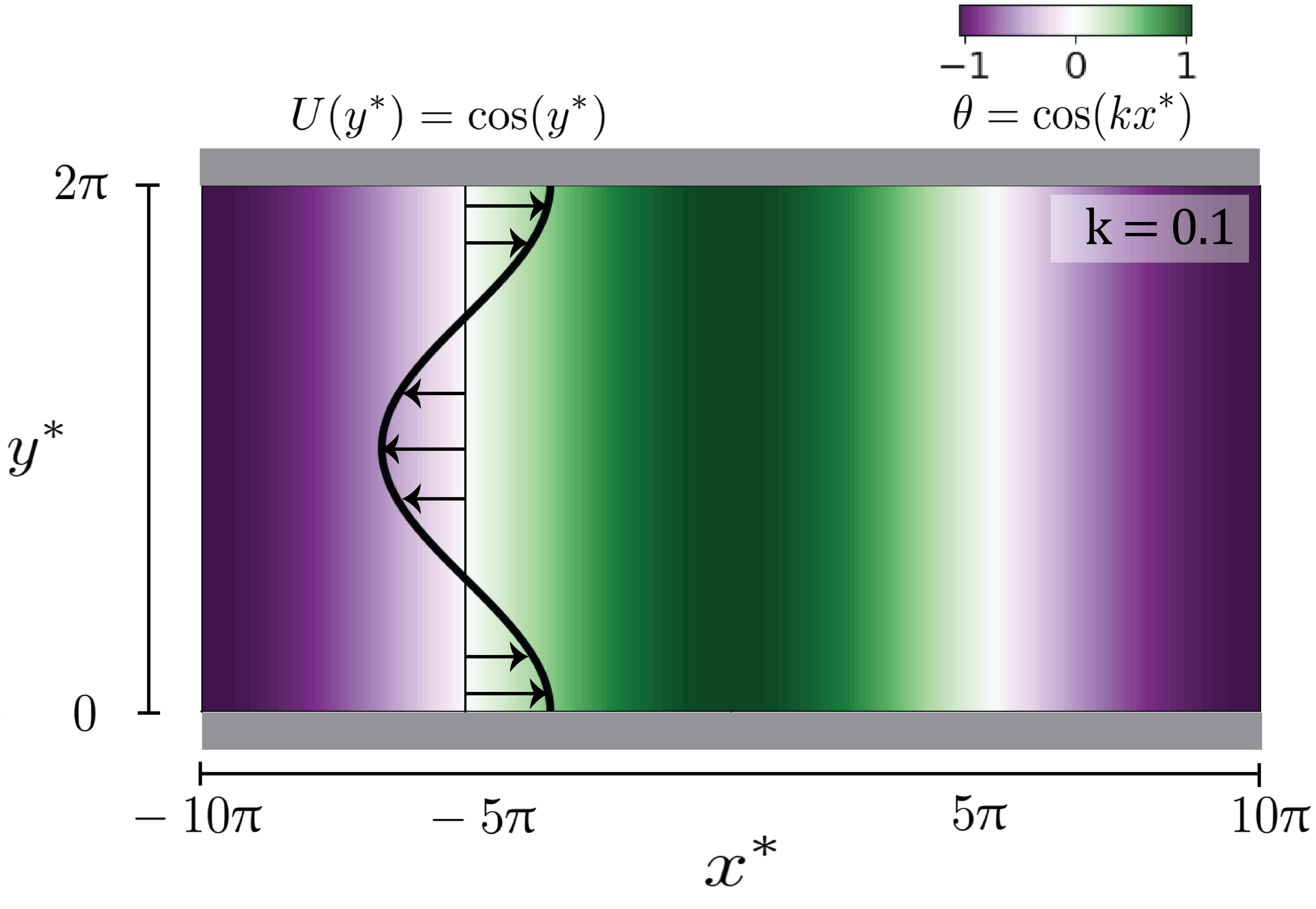}
    \caption{Initial condition of a tracer $\theta(x^*, y^*, 0)$ in a channel that is periodic in $x^*$, and with slippery walls (shaded gray) at $y^*= \{0, \; 2\upi\}$. All variables are non-dimensional. The thick black line and arrows depict the shear flow defined by a single Fourier mode in $y^*$, with wavenumber $\ell=1$. The initial tracer is uniform in the $y$-direction, so that $\Phi(y^*)=1$. The reentrant channel spans $10\upi<x^*< 10\upi$, and so $k\geq0.1$ is the range of non-vanishing wavenumbers that satisfy the periodic boundary condition in $x^*$ for a dimensional length of $\mathcal{L}=10$. A larger choice of $\mathcal{L}$ increases the range of possible $k$-values.}
    \label{fig:IC_diagram}
\end{figure}

Consider the evolution of an inactive passive tracer $\theta=\theta(x, y, t)$ in a re-entrant channel of width $M$. There is a steady, unidirectional, non-divergent horizontal velocity $\mathbf{u}=(U(y), 0)$, where $U(y)=U_{0}\cos{(2\pi y/M)}$, and molecular diffusion with diffusivity coefficient $\kappa$. The advection diffusion equation determines the evolution of the tracer and reads
\begin{equation}\label{eqn}
    \frac{\partial\theta}{\partial t} + U_{0}\cos{\left(\frac{2\pi y}{M}\right)}\frac{\partial\theta}{\partial x} = \kappa\nabla^2\theta  ,
\end{equation}
\begin{equation}\label{general_IC}
    \theta(x, y, 0) = \theta_{0} \cos \left(\frac{2\upi k x}{L} \right)\Phi\left(\frac{2\upi y}{M}\right) ,
\end{equation}
with periodic boundary condition in $x$ and no-flux boundary conditions at $y=\{0, M\}$. The initial condition (\ref{general_IC}) sets the scale of the tracer via the wavenumber $k>0$, regarded as a free parameter. The $x$-length of the channel ($\mathcal{L}$) along with the condition of periodicity in $x$, set the smallest possible value of $k>0$ (and viceversa) as $\mathcal{L}=1/\min{(k)}$ (see Fig. \ref{fig:IC_diagram}). Nonetheless, due to the re entrant boundary condition, the value of $\mathcal{L}$ is physically irrelevant as it can be made a large as desired in order to permit any value of $k>0$ when constructing solutions.

The flow satisfies free-slip boundary conditions and thus, solutions are doubly-periodic. The function $\Phi$ in (\ref{general_IC}) defines the cross-channel ($y$-) structure of the initial condition, with the added restriction for $\Phi$ to be square-integrable so that it can be expressed as a Fourier series. Thus, more complicated $(x, y)$-dependent initial conditions can be constructed as linear combinations of Fourier modes each with different values of $k$.

We non-dimensionlize equation (\ref{eqn}) and the initial condition according to the scaling 
\begin{eqnarray}\label{scaling}
    y = \frac{M y^*}{2\upi},\;\;\;\;\;\;\;\;\; x = \frac{L x^*}{2\upi},\;\;\;\;\;\;\;\;\; t = \frac{M^2 t^*}{4\upi^2\kappa},
\end{eqnarray}
resulting in the non-dimensional advection-diffusion equation
\begin{equation}\label{adv_diff_eqn}
    \epsilon\frac{\partial \theta^*}{\partial t^*}+\cos(y^*)\frac{\partial \theta^*}{\partial x^*} - \frac{1}{\Pen}\frac{\partial^2\theta}{\partial x^{*2}} - \epsilon\frac{\partial^2\theta}{\partial y^{*2}}= 0 ,
 \end{equation}
with initial condition $\theta(x^*, y^*, 0)=\theta_{0}\cos(kx^*) \, \Phi(y^*)$. The non-dimensional domain is then defined as $(-\upi\leq kx^* \leq \upi)\times(0\leq y^*\leq 2\upi)$, and $t^*>0$. There are two non-dimensional parameters in (\ref{adv_diff_eqn}):
\begin{equation}\label{Params}
    \Pen = \frac{U_{0}L}{2\upi \kappa}, \;\;\;\;\;\;\;\;\;\;\;\;\;\;  \epsilon = \frac{t_{a}}{t_{d}} = \frac{L^2}{M^2 \Pen} ,
\end{equation}
where $\Pen$ is the P\'{e}clet number, a measure of the strength of diffusive fluxes compared to advective transport, and $\epsilon$ is the ratio between the advective and diffusive time scales, given by $t_{a}=L/(2\upi U_{0})$ and $t_{d}=M^2/(4\upi^2\kappa)$, respectively. The ratio $M/L$ is referred to as a scale-separation parameter and when considered small it is used to derive asymptotic solutions to (\ref{adv_diff_eqn}) (see \citealt{camassa2010exact}). We make no \textit{a priori} assumption of scale separation and, without loss of generality we set $L=M$, which implies that the only physically relevant length scale is given by the width of the varying flow. Then $\epsilon = \Pen^{-1}$ and thus P\'{e}clet number is the only (free) parameter that arises from the equation (\ref{adv_diff_eqn}). The other independent parameter is the wavenumber of the initial condition $k>0$ and we are interested in all possible choices, so as to compose more complex initial conditions via Fourier synthesis. The range of values $k\ll 1$ is important as it represents scale separation between the (larger) scale of the tracer and the scale of the periodic shear flow  (the scale of the flow is set by a single Fourier mode with wavenumber $\ell\equiv1$;  $U(y)=\cos(\ell y^*)$). Thus, Taylor's effective diffusivity is valid in the limit $\Pen\rightarrow\infty$, $k\ll1$, and $t^* > 1$.

\subsection{Problem Solution}
We seek $x$-periodic solutions\footnote{Note that the stars have been dropped, so henceforth all variables are non-dimensional.} to the homogeneous problem (\ref{adv_diff_eqn}) using the \textit{ansatz}
\begin{equation}\label{prop_soln}
    \theta(x, y , t) =  \Re \left\{A(k, \omega, y) \exp \left[ikx-\omega t\right]\right\} ,
\end{equation}
where the frequency $\omega=\omega(k)$ is a function of $k$ via the dispersion relation of (\ref{adv_diff_eqn}).\footnote{For example, in the 1D heat equation, which is the limit $U_0\rightarrow0, k \rightarrow 0$ in (\ref{eqn}) and (\ref{general_IC}), the dispersion relation is $\omega=k^2$ \citep{deconinck2014method}.} In general, $\omega$ may be complex, with the real part indicating decaying/growing solutions and the imaginary part indicating propagation. This separable solution is motivated by the choice of separable initial condition (\ref{general_IC}). Substituting (\ref{prop_soln}) into (\ref{adv_diff_eqn}) yields the equation for amplitude $A(k, \omega, y)$
\begin{equation}\label{Mathieu_almost}
  \frac{d^2A}{dy^2} + \left[\omega -k^2 - ik\Pen\cos(y) \right]A = 0 ,
\end{equation}
which is, by construction, independent of time. Re-scaling the domain as $\tilde{y}= y/2$, and defining the parameters
\begin{equation}\label{Mathieu_Params}
    a = 4\omega - 4k^2, \;\;\;\;\;\;\;\;\;\; q = 2ik\Pen ,
\end{equation}
converts (\ref{Mathieu_almost}) into the canonical \textit{Mathieu} equation
\begin{equation}\label{Mathieu}
   \frac{d^2A}{d \tilde{y}^{2}}  + \left[a-2q\cos(2\tilde{y})\right]A = 0 .
\end{equation}
In accordace to Floquet theory, solutions to (\ref{Mathieu}) that are $\upi$-periodic in $\tilde{y}$ ($2\upi$-periodic in $y$) and satisfy the no-flux boundary conditions, are the even, cosine-elliptic \textit{Mathieu functions} $ \ce _{2n}(q, \tilde{y})$,  $n=0, 1, 2, \cdots$, also called Mathieu functions of the first kind \citep{McLachlan1947Mathieu, olver2010nist}. Associated with each Mathieu function $\ce_{2n}$ is the eigenvalue $a=a_{2n}(q)$ (also called the characteristic number). Both $\ce_{2n}$ and $a_{2n}$ depend on the parameter $q=2ik\Pen$ \citep{McLachlan1947Mathieu, olver2010nist, arscott2014periodic}. This implies, to our advantage, that the free parameters in the problem collapse into a single compound parameter, Mathieu's canonical parameter $q$, which contains all the relevant physics. Because Mathieu functions converge to Fourier series as $q$ vanishes, $\lim_{q \rightarrow 0} \ce_{2n} (q,\tilde{y}) = \cos(2n\tilde{y})$, $q$ can be considered an eccentricity parameter for each cosine-elliptic function.

\begin{figure}
    \centering
    \includegraphics[width=350pt]{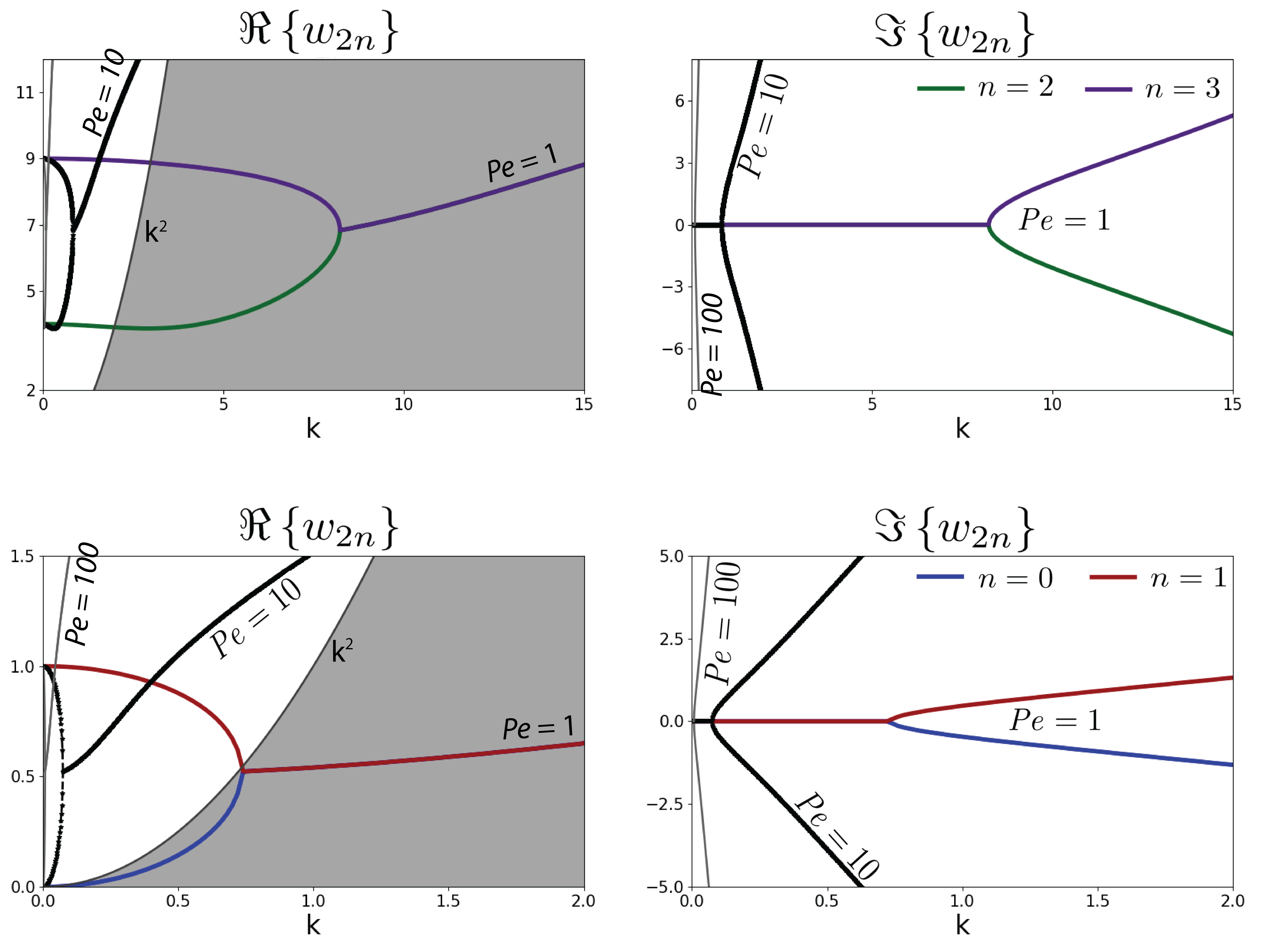}
    \caption{Comparison of the two terms that define the eigenfrequency $w_{2n}$ in (\ref{eigen_freq}) associated with the gravest four Mathieu eigenfunctions $n=0, 1, 2, 3$, and their dependence on wavenumber $k$. The wavenumbers of the exceptional points (EPs) are defined by $|q_{ l }^{EP}| /2\Pen$ where the $q_{ l }^{EP}$ values are fixed (Table \ref{tab:eps}). The EPs shift towards low wavenumbers when $\Pen>1$ and towards high wavenumbers when $\Pen<1$. Colored lines represent $\Pen=1$, with black starred lines $\Pen=10$, and gray lines $\Pen=100$. The shaded region, delimited by the curve $k^2$, shows the range of wavenumbers for which diffusion dominates $\omega_{2n}$, which increases the decay rate of the eigenmodes. Diffusion has no effect on the imaginary part of the eigenvalues.}
    \label{fig:eigs_Pe}
\end{figure}

From (\ref{Mathieu_Params}), the dispersion relation of (\ref{adv_diff_eqn}) is
\begin{equation}\label{eigen_freq}
    \omega_{2n} = \frac{a_{2n}(q)}{4} + k^2 ,
\end{equation}
which sets the time dependence of each eigenfunction $\ce_{2n}$. Specifically, from (\ref{prop_soln}) the time dependence of each $\ce_{2n}$ in $A(k, \omega, y)$ is given by the sum of $k^2$ (1D diffusion) and the eigenvalue $a_{2n}$ (see Fig. \ref{fig:eigs_Pe}). For large enough $k$, $\omega_{2n}$ is dominated by the $k^2$ term and the $\ce_{2n}$ mode decays exponentially with timescale $\tau\sim 1 /\Re\{\omega_{2n}\} \approx 1/k^2$. Otherwise, for small $k$, or large $\Pen$ (or both), $\omega_{2n}$ is dominated by the $a_{2n}$ term and 1D diffusion is negligible. Therefore, the functional dependence of $a_{2n}$ with $q$ (and hence with $k=q / (2 i \Pen)$) is key to understanding the solution. 

The eigenvalues $a_{2n}(q)$ exhibit a transition with increasing $q$ when the parameter $q$ is purely imaginary (as it is here). For $q \lessapprox 1.468i$, the $a_{2n}$ are all real and positive. Therefore, the $\ce_{2n}$ modes all decay exponentially with time. As $q$ grows, successive pairs of eigenvalues merge into complex conjugate pairs (with positive real parts). These mergers happen at {\it Exceptional Points} (EPs), as seen in Fig. \ref{fig:eigs_Pe} \citep{mulholland1929, hunter1981eigenvalues, olver2010nist, ziener2012mathieu}. The EPs for the cosine-elliptic functions $\ce_{2n}$ occur at well known and fixed values of $q$ (Table \ref{tab:eps}), and the $q$ values of the EPs increase with $n$. The gravest eigenvalues $a_{0}, a_{2}$ coalesce (in their real parts) and branch (in their imaginary parts) at $q = q_{0}^{EP} \approx 1.468i$ (Fig. \ref{fig:eigs_Pe}, Table \ref{tab:eps}). The next pair of eigenvalues merges at $q_{1}^{EP} \approx 16.47i$, and so on. Transition of $a_{2n}$ past EPs makes the eigenfrequency $\omega_{2n}$ for that mode complex. This indicates a different physical behavior: the $\ce_{2n}$ mode now propagates along the channel (and decays). The importance of EPs has recently become evident in several areas of mathematical physics \citep{bender1998real, bender1999complex,heiss2004exceptional, heiss2012physics, miri2019exceptional}. Shear dispersion is another example.

\begin{table}
 \begin{center}
  \begin{tabular}{ccc}
$l$ &    $q_{ l }^{EP}$  & Eigenvalue pairs  \\[3pt]
0   &   $1.468i$ & $a_{0}, a_{2}$   \\
1   &  $16.471i$ & $a_{4}, a_{6}$   \\
2   &  $47.806i$ & $a_{8}, a_{10}$  \\
3  &  $95.475i$ & $a_{12}, a_{14}$ \\
  \end{tabular}
  \caption{Locations $q_{ l }^{EP}$ of the first four exceptional points (EPs) for Mathieu's cosine-elliptic eigenfunctions (from \citealt{blanch1969double}). At EPs two successive eigenvalues coalesce (branch) in their real (imaginary) parts with increasing $q$. See Fig. \ref{fig:eigs_Pe}. }{\label{tab:eps}}
 \end{center}
\end{table}

Mathieu functions $\ce_{2n}(q, \tilde{y})$ can be defined by a Fourier series \citep{McLachlan1947Mathieu, arscott2014periodic}. For $\ce _{2n}( q , \tilde{y})$, this is
\begin{equation}\label{ce2n}
     \ce _{2n}( q , \tilde{y}) = \sum_{r=0}^{\infty}A_{2r}^{(2n)}( q ) \cos(2r\tilde{y}) , \;\;\;\;\;\; n=0, 1, 2, \cdots
\end{equation}
For each $n$ and $q$, the eigenvalue $a_{2n} (q) $ and the sequence of Fourier coefficients $A_{2r}^{(2n)}( q )$ are calculated by standard procedures (see Appendix \ref{appA} and the github software repository for this paper). 

\begin{figure}
    \centering
    \includegraphics[width=250pt]{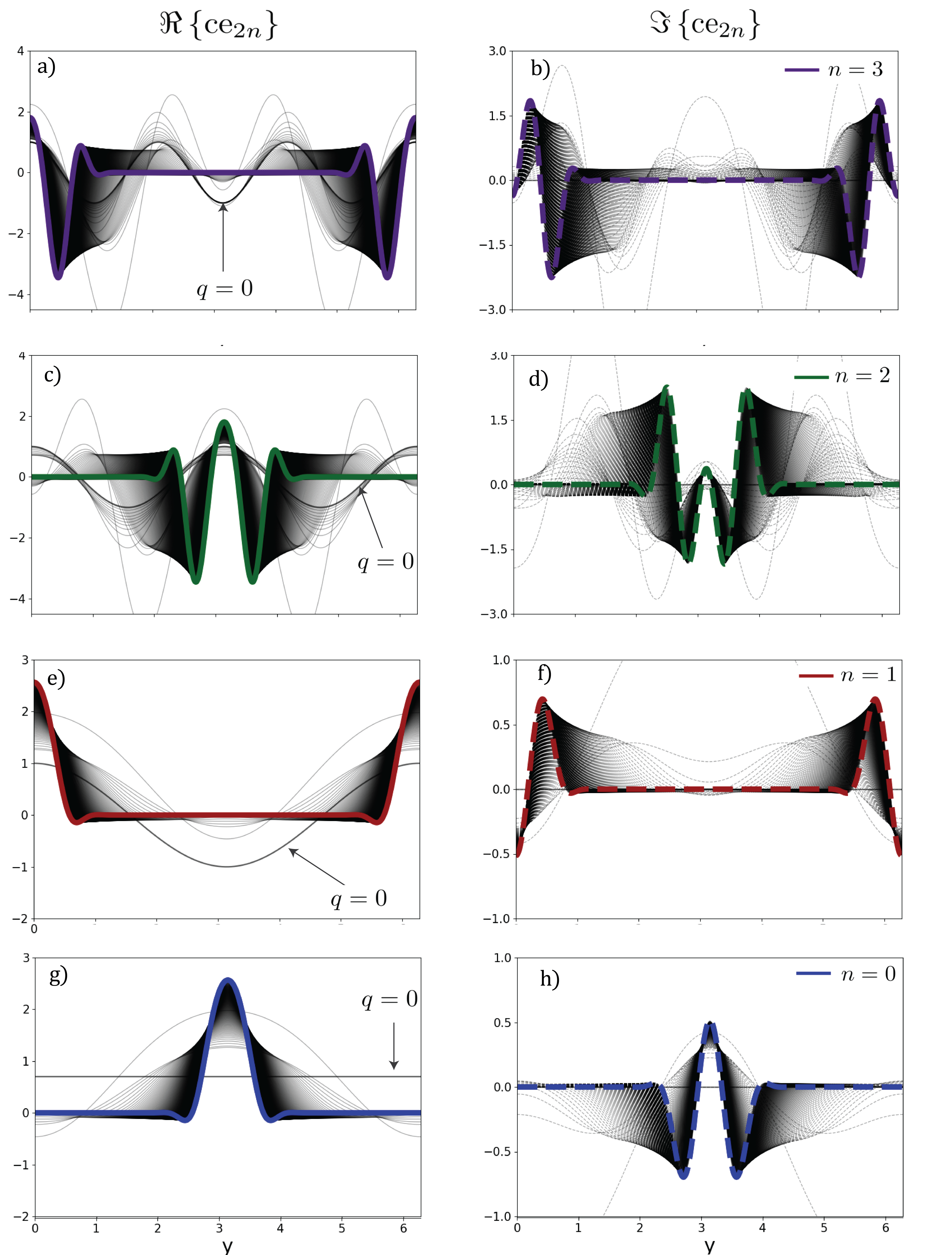}
    \caption{First four Mathieu eigenfunctions $ \ce _{2n}(q, y)$ over a range of the parameter $0< q=2ik\Pen < 900i$, which extends past their EPs (Table \ref{tab:eps}). Colored lines are $\ce_{2n}(q=900i, y)$ and dark gray lines are $\ce_{2n}(q=0, y)$. The $\ce_{2n}$ evaluated at $q=900i$ (colored lines), can equally represent a solution when $k=10$ and $\Pen=45$, or when $k=0.1$ and $\Pen=4500$. Note that the abscissa is $y = 2\tilde{y}$.}
    \label{fig:ce0ce2}
\end{figure}

As $q$ increases past each EP and the two eigenvalues coalesce, the associated eigenfunctions also coalesce to become complex conjugates and for values $q>q_{ l }^{EP}$, these eigenfunctions satisfy a shifted-conjugate symmetry, e.g. $\ce_{2}(q, \tilde{y}) = \ce_{0}^*(q, \tilde{y}-\upi/2)$ where $^*$ denotes the complex conjugate (Fig.~\ref{fig:ce0ce2}). Except at EPs, Mathieu eigenfunctions furnish a complete base, equiconvergent with Fourier series (see \citealt{olver2010nist} Ch. 28). Therefore, they can be used to approximate a $y$-dependent function in the Hilbert space of periodic (square integrable) functions.\footnote{Exactly at the EP, the eigenvalues are repeated and the eigenfunctions no longer span the complete space. Nonetheless, the set of eigenfunctions can be made complete by defining a generalized eigenfunction \citep{brimacombe2020computation}.}

From the general definition (\ref{ce2n}), the exact solution to (\ref{adv_diff_eqn}) is given by a linear combination of even, cosine-elliptic Mathieu functions. This is
\begin{equation}\label{Mathieu_Sol}
   \theta(x, \tilde{y}, t) =\Re \left\{ \sum_{n=0}^{\infty} \alpha_{2n}(q)\: \ce _{2n}(q, \tilde{y})\: \exp \left[ikx - \left(\frac{a_{2n}}{4}+k^2\right)t\right]\right\} .
\end{equation}
The coefficients $\alpha_{2n}$ are those requiring (\ref{Mathieu_Sol}) to satisfy the initial condition
\begin{equation}\label{IC}
   \sum_{n=0}^{\infty}\alpha_{2n} (q) \, \ce _{2n}(q, \tilde{y}) =\Phi(\tilde{y}) .
\end{equation}
Since Mathieu eigenfunctions $ \ce_{2n}(q, \tilde{y})$ are expressed as a Fourier cosine series, and assuming that $\Phi(\tilde{y})$ has a Fourier cosine series, then the individual $\alpha_{2n}$ coefficients are constructed from known Mathieu function identities (\citealt{olver2010nist} Ch. 28). In the case when $\Phi=1$, we have the identity
\begin{align}\label{identity}
    2\sum_{n=0}^{\infty} A^{(2n)}_{0} (q) \, \ce _{2n}(q, \tilde{y}) &= 1 ,
\end{align}
where $A_{0}^{(2n)}$ is the zeroth Fourier coefficient in the definition of each Mathieu function $\ce_{2n}$ in (\ref{ce2n}). Thus from (\ref{IC}) $\alpha_{2n} = 2A_{0}^{(2n)}$, and the exact solution to (\ref{adv_diff_eqn}) with the initial condition $\theta(x, y, 0)=\cos{(kx)}$ is
\begin{equation}\label{Phi1_soln}
    \theta(x, \tilde{y}, t) = \Re \left\{\sum_{n=0}^{\infty} 2\:\;A_{0}^{(2n)}(q)\; \ce _{2n}(q, \tilde{y})\; \exp \left[ikx - \left(\frac{a_{2n}}{4}+k^2\right)t\right] \right\} .
\end{equation}

Similarly, for $\Phi(\tilde{y})= \cos(2p\tilde{y})$, we have the following identity
\begin{align}
\label{cosine_ce_identity}
     \sum_{n=0}^{\infty}A_{2p}^{(2n)} \ce _{2n}(q, \tilde{y}) &= \cos(2 p \tilde{y}) , \;\;\;\;\;\;\;\; r \neq0 ,
\end{align}
where $A_{2p}^{(2n)}$ is the $p$th Fourier coefficient in the definition of Mathieu functions in (\ref{ce2n}). Hence, $\alpha_{2n} = A_{2p}^{(2n)}$ and the exact solution to (\ref{adv_diff_eqn}) with initial condition $\theta(x, y, 0) = \cos(kx)\cos(2p\tilde{y})$ is
\begin{equation}\label{Phi_cos_soln}
    \theta(x, \tilde{y}, t) = \Re \left\{\sum_{n=0}^{\infty} \;A_{2p}^{(2n)}(q)\; \ce _{2n}(q, \tilde{y})\; \exp \left[ikx - \left(\frac{a_{2n}}{4}+k^2\right)t\right] \right\} .
\end{equation}

\begin{figure}
    \centering
    \includegraphics[width=375pt]{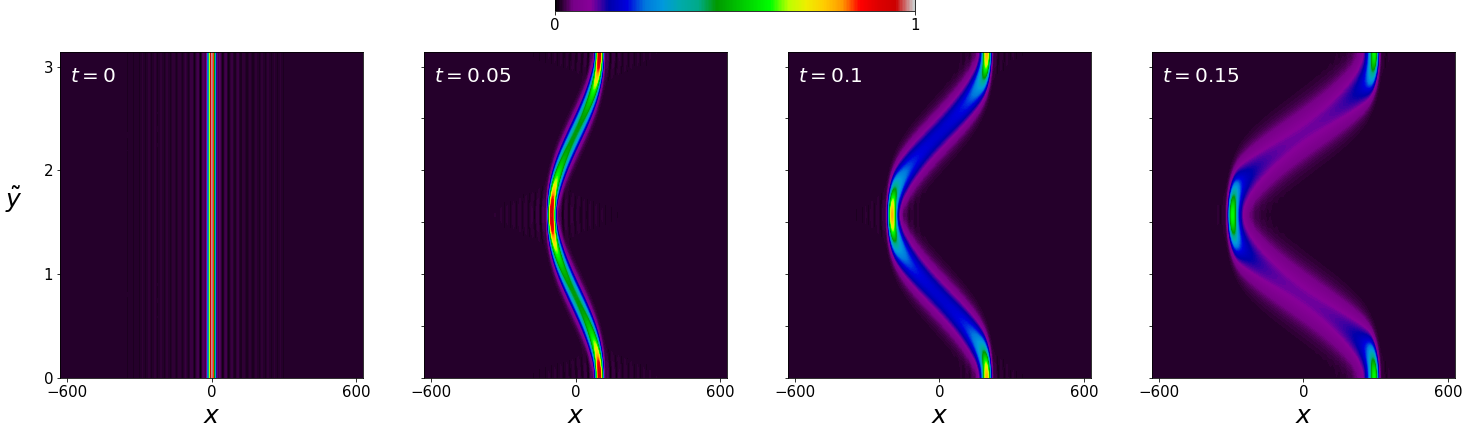}
    \caption{Time evolution of a passive tracer concentrated at $x=0$ defined as a Gaussian in $x$, and $\Phi(\tilde{y})=1$ with $\Pen=2000$, in the presence of a periodic shear flow $(U(\tilde{y}), 0)=(\cos(2\tilde{y}), 0)$ The initial condition results from the linear combination of cosine modes of the form (\ref{Phi1_soln}), with a range of wavenumbers $0<k<0.25$, possible by increasing the domain in $x$ as $-200\upi\leq x \leq 200\upi$. In this initial condition, every mode has a scale much larger than that of the velocity field.}
    \label{fig:gaussian_spread}
\end{figure}

In general, let $\Phi(\tilde{y})$ have a Fourier cosine series of the form
\begin{equation}\label{Phi_cos_series}
    \Phi(\tilde{y}) = \sum_{p=0}^{\infty}\beta_{2p}\cos{(2p \tilde{y})}.
\end{equation}
Then, the general solution to the shear dispersion problem (\ref{eqn}) with initial condition $\theta(x, \tilde{y}, 0) = \cos{(kx)}\, \Phi(\tilde{y}) $ is
\begin{equation}\label{gral_soln}
    \theta(x, \tilde{y}, t) = \Re\left\{\sum_{n=0}^{\infty}\sum_{p=0}^{\infty} (1+\delta_{0p})\, \beta_{2p} \, A_{2p}^{(2n)}(q) \, \ce_{2n}(q, \tilde{y})\exp{\left[ikx-\left(\frac{a_{2n}}{4} + k^2\right)t\right]} \right\} ,
\end{equation}
where $\delta_{0p}$ is the Kronecker delta\footnote{The general solution implies $\alpha_{2n}(q)=\sum_{p=0}^{\infty}(1+\delta_{0p})\beta_{2p}A_{2p}^{(2n)}(q)$ in (\ref{Mathieu_Sol}) and (\ref{IC})}. This solution describes the initial condition consisting of a single Fourier mode in $x$. More general initial conditions can be defined as a linear combination (in $x$) of terms like (\ref{gral_soln}) for different values of $k$. In other words, an initial condition $\theta (x, \tilde{y}, 0) = f(x,\tilde{y})$ is handled by expressing $f(x,\tilde{y})$ as a double Fourier cosine series in $x$ and $y$ and a triple sum that generalizes (\ref{gral_soln}).

\subsection{Example Solution: a Gaussian initial condition}
\label{sec:example}
Although both the purely decaying ($q\rightarrow0$) and propagating regimes ($q\rightarrow\infty$) admit solutions associated with the large P\'{e}clet number limit, only the propagating regime can describe solutions associated with a wide range of wavenumbers in the initial condition. This is well illustrated by a Gaussian initial condition (Fig. \ref{fig:gaussian_spread}). The initial condition is given by a finite sum of cosine-modes in $x$, and the solution is a linear combination of cosine solutions each of the form (\ref{Phi1_soln}), with all modes satisfying $k<0.25$. With a value of $\Pen=2000$, the range in the Mathieu parameter for each cosine-mode is $20i < q <1000i$ and so the gravest mode lies beyond the first two EPs (for $\Pen=2000$ only the wavenumbers $k < 0.0003$ satisfy $q\lessapprox 1.468i$). Similarly, for the longest mode $k=0.25$ to lie before the first EP $q_{0}^{EP}\approx 1.468i$, the P\'{e}clet number must be $\Pen<1.568$. Thus, the limit of large $q$ and large P\'{e}clet number captures both the early and late time behavior of the exact solution.

\section{Closure for Average Shear Dispersion}\label{sec:Closures}

The exact analytical solution to the advection diffusion equation, allows us to explore the exact representation of the averaged equation and thus derive an exact closure equation that is uniform across scales. From the exact closure equation, approximations are derived with particular applicability and utility.

\subsection{Closure Statement}
Consider the family of weighted averages defined by
\begin{equation}\label{gral_average}
     \langle f(\tilde{y}) \rangle _{2m} = \frac{1}{\upi}\int_{0}^{\upi} f(\tilde{y})  \cos (2m \tilde{y})\;d\tilde{y} , \;\;\;\;\;\;\;\; m=0, 1, 2 ,\cdots ,
\end{equation}
where $ \cos(2m\tilde{y})$ represents the weight/filter. 
Applying these operators to an initial condition whose $\tilde{y}$-dependence is a Fourier cosine series, as in (\ref{Phi_cos_series}), gives
\begin{equation}\label{Phi_cos_average}
    \langle \Phi\rangle_{2m} = \beta_{2m}\left(\dfrac{1+\delta_{0m}}{2} \right)
\end{equation}
Thus, the average (\ref{gral_average}) selects the $p=m$ cosine mode in the general initial condition. The average of Mathieu functions (\ref{ce2n}) yields
\begin{equation}\label{ce_ave_2m}
    \langle \ce_{2n}(q, \tilde{y}) \rangle_{2m} = A_{2m}^{(2n)}\left(\dfrac{1+\delta_{0m}}{2}\right)
\end{equation}
Similarly, the average $\langle \ce_{2n} \rangle_{2m}$ selects the single $r = m$ cosine mode in the infinite sum that defines each eigenfunction from (\ref{ce2n}). Note that both in (\ref{Phi_cos_average}) and (\ref{ce_ave_2m}) the averages with $m\neq0$ have a factor of $1/2$ due to the product of cosines when integrating according to (\ref{gral_average}). These definitions allow us to study the time evolution of a single cosine mode in the exact solution.

Applying the average (\ref{gral_average}) to the exact, general solution (\ref{gral_soln}) yields 
\begin{equation}\label{mean_theta}
     \langle \theta  \rangle _{2m} =  \Re \left\{\sum_{n=0}^{\infty}\sum_{p=0}^{\infty}(1+\delta_{0p})\beta_{2p}A_{2p}^{(2n)}A_{2m}^{(2n)}\left((1+\delta_{0m})/2 \right) \exp \left[ikx - \left(\frac{a_{2n}}{4} + k^2\right)t \right] \right\} .
\end{equation}
The initial condition $ \langle \theta(x,y,0) \rangle _{2m}= \langle \cos(k x)\; \Phi (y) \rangle_{2m} =  \cos(k x)\;\beta _{2m}\left((1+\delta_{0m})/2 \right)$ is guaranteed for all choices of $m$ as long as $\Phi$ is defined as in (\ref{Phi_cos_series}) due to the orthogonality of the $A_{2p}^{(2n)}$ coefficients that defined Mathieu functions (see Appendix \ref{appA} and (\ref{ortho})).

Applying the average (\ref{gral_average}) to the advection-diffusion equation (\ref{adv_diff_eqn}) yields
\begin{equation}\label{mean_eqn}
    \frac{\partial \langle \theta \rangle _{2m}}{\partial t} +  \Pen\frac{\partial\left( \langle \theta \cos(2\tilde{y}) \rangle _{2m}\right)}{\partial x} = \frac{\partial^2 \langle \theta \rangle _{2m}}{\partial x^2}
    - 4m^2 \langle \theta \rangle_{2m}
\end{equation}
for all $m=0, 1, 2, \cdots$. Interest lies in the \textit{closure} of the advection term, namely, the second term on the left hand side of (\ref{mean_eqn}). We aim to write it exactly as a differential operator acting on the mean tracer field and to approximate it, so as to render a simpler compact form.

\subsection{Closure Solution}
The expression of interest is
\begin{eqnarray}\label{closure_reln}
\frac{\partial\left( \langle \theta \cos(2\tilde{y})\rangle _{2m}\right)}{\partial x} \equiv  \Lambda _{2m}\left[ \langle \theta \rangle _{2m}\right] ,
\end{eqnarray}
where $\Lambda _{2m}$ is the closure operator acting on the mean tracer $ \langle \theta \rangle _{2m}$. The left hand side of (\ref{closure_reln}) is given exactly by
\begin{equation}\label{lhs_closure}
    \Re\left\{ik\exp\left(ikx\right)\sum_{n=0}^{\infty}\sum_{p=0}^{\infty}(1+\delta_{p0})A_{2p}^{(2n)}\langle\ce_{2n}\cos{(2\tilde{y})}\rangle_{2m}\exp\left[-\left(\dfrac{a_{2n}}{4}+k^2\right)t \right] \right\}
\end{equation}
Since the exact averaged solution (\ref{mean_theta}) can be written as the infinite sum $ \langle \theta \rangle _{2m} =  \Re \left\{\sum_{n=0}^{\infty} \langle\theta^{(2n)}\rangle_{2m}\right\}$, then (\ref{lhs_closure}) implies that the closure equation (\ref{closure_reln}) is written for each $\langle\theta^{(2n)}\rangle_{2m}$ as follows
\begin{equation}\label{closure_modal}
    ik\langle\ce_{2n}(q, \tilde{y}) \cos(2\tilde{y})\rangle_{2m} = \Lambda^{(2n)}_{2m} A_{2m}^{(2n)}\left(\dfrac{1+\delta_{0m}}{2}\right)
\end{equation}
Thus the averaged, unclosed advection-diffusion equation (\ref{mean_eqn}) can be written as an autonomous, closed system of equations for each $n$-mode, which in Fourier space reads like
\begin{equation}\label{modal_closure_eqn}
    \frac{\partial \langle\theta^{(2n)}\rangle_{2m}}{\partial t} = -\left(k^2 + 4m^2 + \Pen\Lambda_{2m}^{(2n)} \right)\langle\theta^{(2n)}\rangle_{2m} .
\end{equation}
The $n$-modes are subject to the initial condition
\begin{equation}
 \Re \left\{\sum_{n=0}^{\infty}\langle\theta^{(2n)}\rangle_{2m}(x, t=0) \right\}= \beta_{2m}\left(\dfrac{1+\delta_{0m}}{2}\right)\cos (k x)    .
\end{equation}
We  thus proceed by first computing the left hand side of (\ref{closure_modal}), which contains all the $\tilde{y}$ dependence in $\theta$. This is
\begin{align}\label{partial_sum}
    2 \langle \cos(2\tilde{y}) \ce_{2n}(q, \tilde{y}) \rangle _{2m} &= 2\sum_{r=0}^{\infty}A_{2r}^{(2n)} \langle \cos(2\tilde{y})\cos(2r\tilde{y}) \rangle _{2m}\nonumber , \\
    &= \phantom{2} \sum_{r=0}^{\infty}A_{2r}^{(2n)} \langle  \cos \left( 2\left[r-1\right]\tilde{y} \right) + \cos \left( 2\left[r+1\right]\tilde{y} \right) \rangle _{2m} ,
\end{align}
where the Mathieu function definition (\ref{ce2n}) is used. For $m=0$ in the weighted average (\ref{gral_average}) (arithmetic cross-channel mean), the only term in the infinite sum that survives is the $r=1$ Fourier coefficient, for each $n=0, 1, 2, \cdots$. Thus, when $m=0$,
\begin{equation}
     \langle \cos(2\tilde{y}) \ce_{2n}(q, \tilde{y}) \rangle _{0} = \frac{A_{2}^{(2n)}}{2} .
\end{equation}

For $m>0$, the $\langle \cdot \rangle_{2m}$ in (\ref{partial_sum}) involves the two cosine terms multiplying the weight $ \cos (2m\tilde{y})$. They are rewritten as
\begin{align}\label{first_sum}
    2 \cos \left( 2 \left[r \pm 1 \right]\tilde{y}\right) \cos (2m\tilde{y}) & = \cos \left( 2 \left[ r+m \pm 1 \right] \tilde{y}\right) + \cos \left( 2 \left[ r-m \pm 1 \right] \tilde{y}\right) ,
\end{align}
again using trigonometric identities.
The non-vanishing terms in the average are those for which the arguments of either cosine on the right hand side of (\ref{first_sum}) vanish. Consider first $m=1$ after which we generalize to all $m \ge 2$.
When $m=1$, only $r=0$ contributes to (\ref{first_sum}) with the plus sign, whereas both $r=0$ and $r=2$ contribute with the minus sign. Therefore, when $m=1$
\begin{equation}\label{corr_m1}
    \langle\cos(2\tilde{y})\ce_{2n} (q, \tilde{y}) \rangle_{2} = \frac{2A_{0}^{(2n)} + A_{4}^{(2n)}}{4} .
\end{equation}
When $m \ge 2$, only $r=m-1$ contributes to (\ref{first_sum}) with the plus sign, and only $r=m+1$ contributes with the minus sign. Therefore, when $m \ge 2$,
\begin{equation}\label{correlation_gral}
     \langle  \cos(2\tilde{y}) \; \ce _{2n} (q, \tilde{y}) \rangle _{2m} = \frac{A_{2(m-1)}^{(2n)} + A_{2(m+1)}^{(2n)}}{4} .
\end{equation}

With these, we proceed to calculate the exact closure operator in (\ref{closure_modal}) for each choice of $m$ and $n$ as follows
\begin{align}
   ik A_{2}^{(2n)} &=  2\Lambda^{(2n)}_{0} A_{0}^{(2n)} \text{~~~~~} m=0 , \\
   ik \left[ 2A_{0}^{(2n)} + A_{4}^{(2n)} \right] &= 2 \Lambda_{2}^{(2n)} A_{2}^{(2n)} \text{~~~~~} m=1 , \\
   ik \left[A_{2(m-1)}^{(2n)} + A_{2(m+1)}^{(2n)} \right] &= 2 \Lambda^{(2n)}_{2m} A_{2m}^{(2n)} \text{~~~~~} m \ge 2 .
\end{align}
Or, written as a recurrence for $m$
\begin{align}\label{recurrence_closure}
    \Lambda^{(2n)}_{0}A_{0}^{(2n)} - \left(\frac{ik}{2}\right)A_{2}^{(2n)} &= 0 \text{~~~~~} m=0, \nonumber\\
    \Lambda_{2}^{(2n)}A_{2}^{(2n)} - \left(\frac{ik}{2}\right)\left[2A_{0}^{(2n)} + A_{4}^{(2n)}\right] & = 0 \text{~~~~~} m=1, \nonumber\\
    \Lambda^{(2n)}_{2m}A_{2m}^{(2n)} - \left(\frac{ik}{2}\right)\left[A_{2(m-1)}^{(2n)} + A_{2(m+1)}^{(2n)}\right] & = 0 \text{~~~~~}  m \ge 2 .
\end{align}
Remarkably, this recurrence relation matches term by term the recurrence associated with Mathieu's eigenvalue system (see Appendix \ref{appA}, (\ref{recurrence})). Hence, for each choice of mode $m$ in the averaging filter (\ref{gral_average}), we arrive at a family of exact closures for each eigenfunction $ \ce_{2n}$. Each closure operator is proportional to the eigenvalue $a_{2n}$ of the associated eigenfunction:
\begin{equation}\label{exact_closure}
   \Lambda_{2m}^{(2n)} = \dfrac{a_{2n} - 4m^2}{4\Pen}
\end{equation}
The closure above applies for all $m=0, 1, 2, \cdots$.

It is instructive to write the closed equation (\ref{modal_closure_eqn}), say for $m=0$, in physical space and in terms of dimensional variables. From (\ref{scaling}) and (\ref{Params}) the closed dimensional equation is 
\begin{equation}\label{dimensional_closed_eqn}
    \dfrac{\partial\langle\theta^{(2n)}\rangle_{0}}{\partial t} = \kappa\frac{\partial^2\langle\theta^{(2n)}\rangle_{0}}{\partial x^2} - \frac{4\upi^2\kappa\Pen}{M^2}\Lambda^{(2n)}_{0}\left[\langle\theta^{(2n)}\rangle_{0}\right] ,
\end{equation}
where $\Lambda^{(2n)}_{0}$ is an operator acting on the averaged mode $\langle\theta^{(2n)}\rangle_{0}$. 

\subsection{Asymptotic Closure for Small and Large Canonical Parameters}
In (\ref{exact_closure}) we found that the exact closure for each eigenfunction is proportional to the eigenvalue $a_{2n}(q)$, which is a function of the canonical parameter $q=2ik\Pen$. To express the closures analytically, and to better understand the closures across the two regimes of the solution, consider the asymptotic limits of small and large $q$.

For $q\rightarrow0$ the eigenvalues $a_{2n}$ and therefore the closures $\Lambda^{(2n)}_{2m}$ can be approximated asymptotically as an even power series in $q$ (see Appendix \ref{appB}). The first term in the expansion of the closure operator for the first four eigenfunctions for the cross-channel arithmetic mean $m=0$ reads
\begin{eqnarray}\label{closure_smallq}
    \Lambda_{0}^{(0)} \sim & \phantom{4 \epsilon - } \frac{1}{2}k^2\Pen  + O(k^4\Pen^3), \nonumber\\
    \Lambda_{0}^{(2)} \sim & \phantom{4} \frac{1}{\Pen} - \frac{5}{16}k^{2}\Pen  + O(k^4\Pen^3), \nonumber\\
    \Lambda_{0}^{(4)} \sim & \frac{4}{\Pen} - \frac{1}{30}k^{2}\Pen  + O(k^4\Pen^3), \nonumber\\
    \Lambda_{0}^{(6)} \sim & \frac{9}{\Pen} - \frac{1}{70}k^{2}\Pen  + O(k^4\Pen^3) .
\end{eqnarray}
The presence of the $k^2\Pen$ terms suggests a second derivative that enhances the along-stream diffusion in the original advection-diffusion equation (\ref{adv_diff_eqn}). This follows by connecting Fourier space and physical space with $ik \xleftrightarrow{}\partial_{x}$. Thus, in the $q\rightarrow0$ limit the closure $\Lambda_{0}^{(0)}$, written dimensionally following (\ref{dimensional_closed_eqn}), enhances the molecular diffusion as 
\begin{equation}\label{eff_diff_oper}
    \kappa\left(1+\frac{1}{2}\Pen^2\right)\frac{\partial^2 \langle \theta^{(0)} \rangle _{0}}{\partial x^2} .
\end{equation}
Using (\ref{Params}) this term is
\begin{equation}\label{effective_diffu}
    \frac{\kappa \Pen^2}{2} = \frac{M^2 U_{0}^2}{8\upi^2 \kappa} ,
\end{equation}
which is Taylor's effective diffusivity \citep{taylor1953dispersion}. This approximation is valid in the limit $|q|=2k\Pen\rightarrow 0$, which is attained either by large $\Pen$ under the scale-separation condition $2k \ll 1/\Pen$, or by small P\'{e}clet number for finite $k>0$. This latter limit is characterized by strongly diffusive, weakly advective flows.
 
The closure in (\ref{eff_diff_oper}) is for the gravest, slowest-decaying mode ($n=0$). A similar operator applies to the other modes $n>0$, with a term proportional to $\Pen^2$. Higher order terms in the expansion (see Appendix \ref{appB}, equation (\ref{small_q_as})) provide further corrections to describe the behavior of the exact closures (\ref{exact_closure}) in the small $q$ limit.

In the asymptotic limit $q\rightarrow\infty$, the eigenfrequencies $\omega_{2n}$ are complex, with the imaginary part given by $\Im\{a_{2n}\}/4$ (see Appendix \ref{appB}, (\ref{large_q_as}) and (\ref{large_q_as_46})). The group and phase velocities for the first eigenmode pair are
\begin{eqnarray}\label{cg_cp_vels}
    \dfrac{\partial\Im\{\omega_{0,2}\}}{\partial k} \sim&\pm \dfrac{1}{4}\sqrt{\dfrac{\Pen}{k}} \mp \Pen ,\\ \frac{\Im\{\omega_{0,2}\}}{k} \sim& \pm\dfrac{1}{2}\sqrt{\dfrac{\Pen}{k}} \mp \Pen,
\end{eqnarray}
They differ only by a constant pre-factor of two, independent of $k$ and $\Pen$, signaling dispersive wave behavior. However, as $k\rightarrow\infty$, the propagating modes become non-dispersive, in the kinematic sense. Such behavior is typical of all eigenmodes in the $q\rightarrow\infty$ regime.

For the first four eigenmodes, $q \rightarrow \infty$, and $m=0$, the closures are
\begin{eqnarray}\label{closure_infq}
    \Lambda^{(0, 2)}_{(0)} \sim & \dfrac{1}{2}\sqrt{\dfrac{k}{\Pen}} -\dfrac{1}{16\Pen} \pm i\left[ \dfrac{1}{2}\sqrt{\dfrac{k}{\Pen}} - k\right] , \nonumber\\
    \Lambda^{(4, 6)}_{(0)} \sim & \dfrac{5}{2}\sqrt{\dfrac{k}{\Pen}} - \dfrac{13}{16\Pen} \pm i\left[ \dfrac{5}{2}\sqrt{\dfrac{k}{\Pen}} - k\right] ,
\end{eqnarray}
where we have paired the closures of coalesced eigenvalues for simplicity of expression. In physical space, the closure $\Lambda_{0}^{(0)}$ (with plus sign in imaginary component), can be written as
\begin{equation}\label{closed_eqn_large_Pe_q}
   - U_{0}\frac{\partial\langle\theta^{(0)}\rangle_{0}}{\partial x} - \frac{\upi\sqrt{2\kappa U_{0}}}{M}\frac{\partial^{1/2}\langle\theta^{(0)}\rangle_{0}}{\partial x^{1/2}}.
\end{equation}
where we have used $(1+i)/\sqrt{2} = \sqrt{i}$ and $(ik)^{\alpha}\hat{f}(k) \xleftrightarrow{} \partial^{\alpha}f/\partial x^{\alpha}$ when writing the derivative of fractional order with $\alpha=1/2$ (see \citealt{podlubny1998fractional, tseng2000computation}). The first term represents linear advection, associated with the (kinematically) non-dispersive (i.e. $\pm ik$) term in (\ref{closure_infq}). The second term in (\ref{closed_eqn_large_Pe_q}) is referred to in the literature as a semi-derivative or half-derivative operator (see \citealt{oldham1974fractional} Ch. 3). It captures both the real part of the closure associated with the decay rate of the mode, and the imaginary part associated with the (kinematically) dispersive behavior of the modes (e.g. see \ref{cg_cp_vels}).

\subsection{Approximation to Closure Operator for Arbitrary Wavenumber and P\'{e}clet number}
\begin{figure}
    \centering
    \includegraphics[width=200pt]{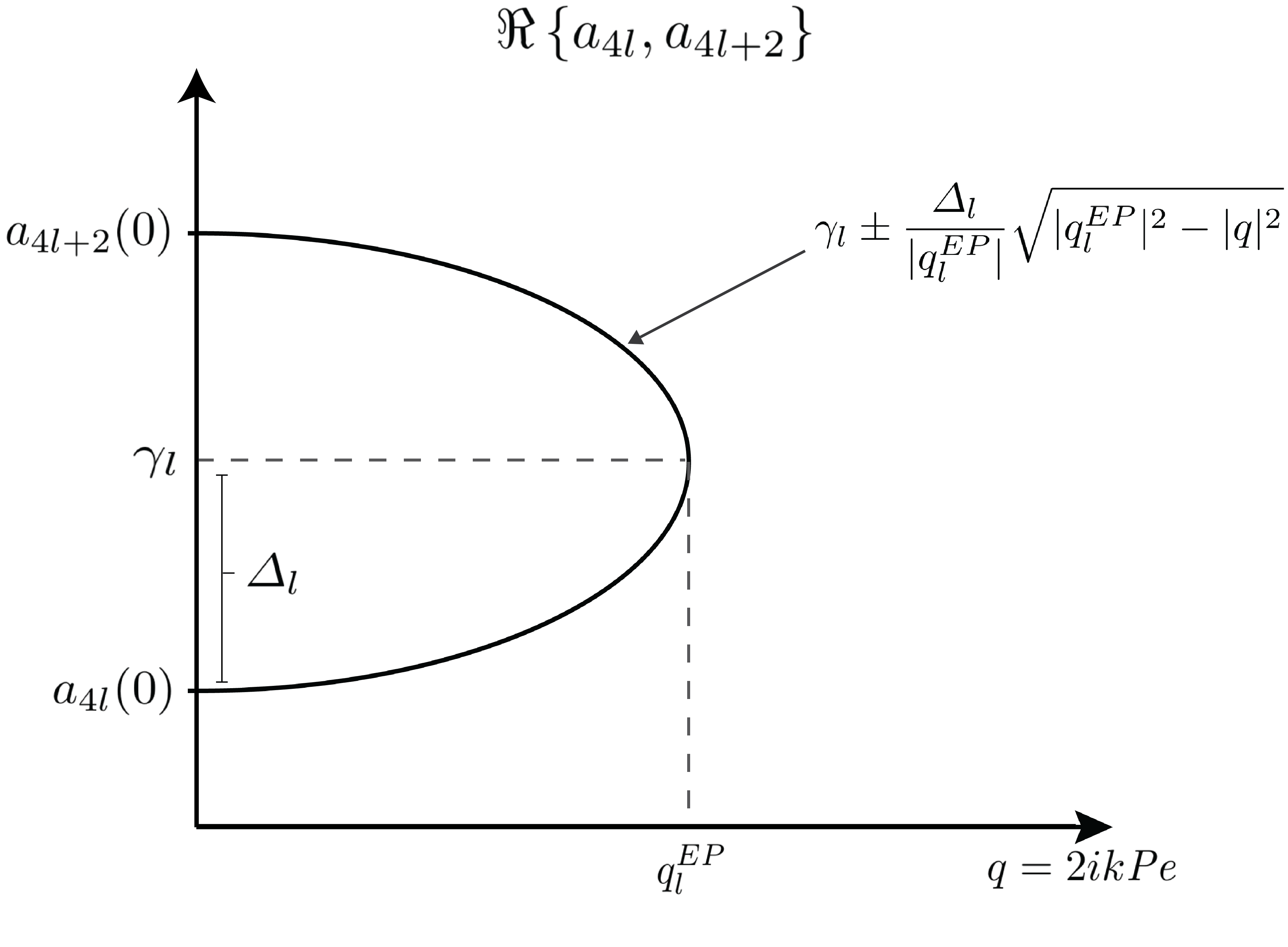}
    \caption{Quadratic approximation to the coalescing eigenvalue pair using a half ellipse. With the known value of $a_{2n}=(2n)^2$ at $q=0$ for all $n=0, 1, 2, \cdots$, we have $\Delta_{ l } = (4  l  + 2)^2 - (4  l )^2 =  8  l +2$ and $\gamma_{ l }=(4  l )^2 + \Delta_{ l } = 16  l ^2+8 l +2$. These parameters are independent of $\Pen$ and $k$. For $q>q_{ l }^{EP}$, the square root becomes purely imaginary and, as $q\rightarrow\infty$, it approximates the linear asymptotic behavior of the coalesced eigenvalues, namely $\sim iq$ at large $q$ (see Appendix \ref{appB}).}
    \label{fig:quadratic_diagram}
\end{figure}

The asymptotic approaches above fail to provide a closure that is continuous across the EPs in wavenumber space, as they only apply for values of $q$ far from EPs. To derive a continuous closure operator, we approximate the eigenvalues as coalescing pairs and divide the parameter dependence into three regions based on the behavior of eigenmodes: I) A region up to the EP that captures the merging of the eigenvalues, $q \le q_{ l }^{EP}$. II) An intermediate matching region for $q_{ l }^{EP} < q \le 2 q_{ l }^{EP}$. And III) a region that matches the asymptotic behavior of eigenvalues $q > 2 q_{ l }^{EP}$. The parameter dependence of the eigenvalues in regions I and II is captured by a \textit{quadratic} approximation to the coalescing pair based on the equation describing a half ellipse\footnote{see also \citealt{ziener2012mathieu} for an alternative quadratic approximation that applies to the lowermost two eigenvalues.}, shown in Figure \ref{fig:quadratic_diagram}.

To capture the coalescing eigenvalue pair that occurs in region I, two pieces of information are needed: 1) The location of each eigenvalue at $q=0$, which is given exactly by $a_{2n} = (2n)^2$, and 2) the location of the EPs on the $q$-axis (i.e. $q=q_{\ell}^{EP}$), which can all be easily calculated independent of any (tunable) physical parameter \citep{blanch1969double, hunter1981eigenvalues}. The quadratic approximation to each pair is therefore
\begin{equation}\label{ellipse_eqn}
    \Gamma_{ l } = \gamma_{ l } \pm \frac{\Delta_{ l }}{|q_{ l }^{EP}|}\sqrt{|q_{ l }^{EP}|^2 - 4k^2\Pen^2} ,
\end{equation}
where $\Gamma_{ l } \approx \left\{a_{4  l }, a_{4  l +2} \right\}$ approximates the eigenvalue pair, $q_{ l }^{EP}$ is the location of the EP on the $q$-axis (Table \ref{tab:eps}), and $\Delta_{ l } = 8  l  + 2$ and $\gamma_{ l } = 16  l ^2 + 8  l  + 2$ are known constants that match the $q$-dependence of the eigenvalue pair (see Fig. \ref{fig:quadratic_diagram} for a geometrical interpretation of the terms in the quadratic approximation).

Approximation captures the coalescing of eigenvalue pairs in region I and provides a continuous transition across (the exact) EPs in wavenumber space from purely real (decaying) to purely imaginary (propagating) in region II. For values of $q>q_{\ell}^{EP}$ the real part of the approximation if a constant ($\gamma_{\ell}$), in disagreement with the behavior of the eigenvalues (see Fig. \ref{fig:eigs_Pe}). Furthermore, the approximation (\ref{ellipse_eqn}) underestimates the amplitude of imaginary part of the coalesced eigenvalues at large $q$ (also beyond the EP), as it approximates the modes as (kinematically) non-dispersive waves. Thus, in region II we add a term based on the asymptotic behaviour of the eigenvalues at large $q$, and region III matches the asymptotic behavior of the coalesced eigenvalue pair at large q.

\begin{figure}
    \centering
    \includegraphics[width=350pt]{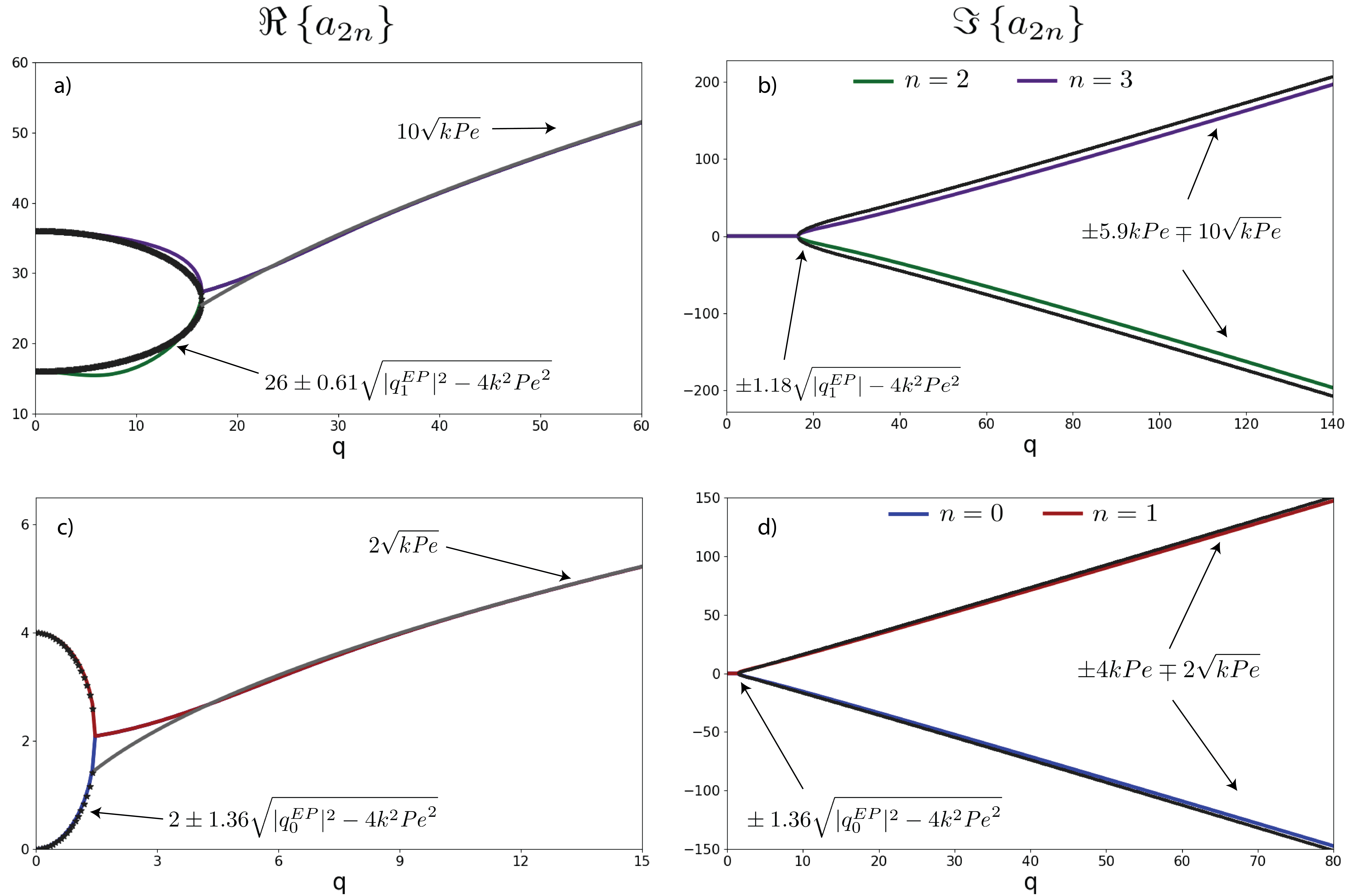}
    \caption{Quadratic approximation across EPs to eigenvalue pairs $\{a_{0},a_{2}\}$ (red and blue lines in (c) and (d)), and $\{a_{4},a_{6}\}$ (green and purple lines in (a) and (b)) in the three regions as described in (\ref{quadratic_operator}). The parameters associated with the quadratic approximation (\ref{ellipse_eqn}) are $\Delta_{0}=2$, $\Delta_{1}=10$ and $\gamma_{0}=2$ and $\gamma_{2}=26$. Depicted in the figures are the closure equations in regions II and III.}
    \label{fig:eigs_closure}
\end{figure}

With the above corrections, the approximation to the exact closure operator (\ref{exact_closure}) that remains continuous in wavenumber space across EPs, captures the small and large $q$-limits, and is valid for all $t>0$ is
\begin{equation}\label{quadratic_operator}
\tilde{\Lambda_{0}}^{(l)} = 
\begin{cases}
    \dfrac{\gamma_{ l }}{4\Pen} \pm \dfrac{\Delta_{ l }}{4|q_{ l }^{EP}|\Pen}\sqrt{|q_{ l }^{EP}|^2 - 4k^2\Pen^2} & \text{if $2k\Pen\leq |q_{ l }^{EP}|$,}\\
   \beta_{ l }\sqrt{\dfrac{k}{\Pen}} \pm\dfrac{\alpha_{\ell}\Delta_{ l }}{4|q_{ l }^{EP}|\Pen}\sqrt{|q_{ l }^{EP}|^2 - 4k^2\Pen^2}  & \text{if $|q_{ l }^{EP}|<2k\Pen\leq 2|q_{ l }^{EP}|$,}\\
    \beta_{ l }\sqrt{\dfrac{k}{\Pen}} \pm i\left(\beta_{ l }\sqrt{\dfrac{k}{\Pen}} - \eta_l k\right) & \text{if $2k\Pen > 2|q_{ l }^{EP}|$.}
    \end{cases}
\end{equation}
In the above approximation, region II is delimited by the range $q^{EP}_{\ell} < q< 2q^{EP}_{\ell}$, although a different upper limit could have been chosen. All of $\alpha_{\ell}, \beta_{\ell}$ and $\eta_{\ell}$ are constants that match the asymptotic limits derived in (\ref{closure_infq}) in a way that assures  the closures are continuous across regions, although a different definition of upper limit in region II would affect these values. Fig. \ref{fig:eigs_closure} shows the closure approximation to the first two eigenvalue pairs $ l =0$, $ l =1$. 

The closure associated with the square-root term in regions I and II in (\ref{quadratic_operator}) that captures the coalescing of eigenvalues retains the limits of large P\'{e}clet number. This is, applying L'Hopital's rule
\begin{eqnarray}\label{operator_limits}
\lim_{Pe\rightarrow\infty} \dfrac{\sqrt{|q_{ l }^{EP}| - 4k^2\Pen^2}}{\Pen} \approx & - k^2\Pen .
    \label{operator_limits2}
\end{eqnarray}
The limit is proportional to the \textit{effective diffusivity} derived by G. I. Taylor as demonstrated in (\ref{closure_smallq}), (\ref{eff_diff_oper}) and (\ref{effective_diffu}).  Furthermore, the argument in the square-root in (\ref{operator_limits2}) must remain positive in order to represent a purely diffusive process. This implies $4k^2\Pen \lessapprox 1.468$ (as $q_{0}^{EP}\approx 1.468i$), which is a statement of strong separation of scales in the large P\'{e}clet number limit, in line with homogenization theory \citep{majda1999simplified, haynes2014dispersion}.

The operator in (\ref{quadratic_operator}) that captures the eigenvalue dependence on $q=2ik\Pen$ across EPs, i.e. across regions I and II, is a fractional order operator in physical space. It can be written
\begin{equation}\label{first_frac_oper}
    \dfrac{1}{\Pen}\sqrt{|q_{ l }^{EP}|\mathcal{I} + 4\Pen\dfrac{\partial^2}{\partial x^2}}
\end{equation}
where $\mathcal{I}$ is an identity operator (when discretized, it is an identity matrix). The operators that capture the asymptotic limit for large $q$-values, i.e. those is region III
in (\ref{quadratic_operator}) are given by the square-root derivatives of the form (\ref{closed_eqn_large_Pe_q}).

\section{Discussion}
\label{sec:discussion}
The results obtained in this study are based on the exact solution to the problem of tracer dispersion by a periodic shear flow for all $t>0$. Our solution method has no \textit{a priori} assumption of separation of scales, but the periodic shear flow is time-independent and so the velocity field does not evolve according to the Navier Stokes equation, a common practice in studies of the advection-diffusion equation \citep{majda1999simplified}. As a result, the closures derived are kinematic, and not dynamic. We neglect any external forcing and reaction terms, but they are readily incorporated.
Our method of solution relies on an \textit{ansatz} (\ref{prop_soln}) that reduces the original advection-diffusion equation into a periodic, second order ordinary differential equation for the amplitude. When the velocity field is defined by a single cosine mode, as in this study, the amplitude equation reduces to the Mathieu equation (\ref{Mathieu}), which depends on a single parameter $q$. The general case specifies the velocity field by a sum of (infinitely) many modes. The resulting amplitude equation is Hill's equation, from which Mathieu's equation is the simplest case, and both are subject to Floquet theory (\citealt{olver2010nist} Ch. 28). Unlike Mathieu's equation, Hill's equation can depend on multiple parameters, which increases the mathematical complexity of the system. Nonetheless, Hill's equation generalizes the problem studied here and thus represents a promising avenue of research. Solutions experience a transition in wavenumber space across Exceptional Points, which are associated with the merger of eigenvalue pairs. Exactly at at the EPs, one eigenfunction is no longer independent, and so the eigenfunctions no longer form a complete set in the space of square-integrable functions. Nonetheless, a generalized eigenfunction can be constructed that completes the set, which solves the shear dispersion problem for all $q$, and thus all possible combinations of $k$ and $\Pen$  (see \citealt{brimacombe2020computation}).
The choice of boundary conditions (in y) determines the eigenfunctions of the homogeneous solution in the initial value problem. We consider no-flux boundary conditions, and so $\ce_{2n}$ determines the solution. As a result, $\Phi$ in the initial condition must have a series expansion in terms of $\ce_{2n}$, i.e. a cosine Fourier series. In the case of Dirichlet boundary conditions, the eigenfunctions are the odd, $\upi$-periodic sine-elliptic Mathieu functions $\se_{2n+2}$, with eigenvalues denoted by $b_{2n+2}$. Even though $\ce_{2n}$ and $\se_{2n+2}$ do not share the same eigenvalues ($a_{2n}\neq b_{2n+2}$), the asymptotic limits at small and large $q$ of the two have the same dependence with $q$, with only slight differences in the coefficients (see \citealt{olver2010nist}). Thus, the closures associated with the choice of Dirichlet boundary conditions are very similar in both Fourier and physical space, as those derived in this paper. Furthermore, because the canonical parameter $q=2ik\Pen$ is insensitive to the choice of boundary conditions, the eigenvalue system (in Appendix \ref{appA}) remains non-Hermitian, and EPs still occur, albeit at different $q$ values.
The P\'{e}clet number dependence of our derived closures elucidates tracer transport by jets. \citet{smith2005tracer} postulated a predictive theory associated with tracer transport by coherent jets in $\beta$-plane turbulence, in which the across-jet transport is described by mixing-length theory and the along-jet transport is controlled by shear dispersion, with a background diffusivity determined by the across-jet mixing. Such theory, valid in the limits of homogenization ($\Pen$ large but $|q|\ll1$), implies that changes to the background diffusivity have an inverse effect on the along-jet diffusive transport. In our derived closure in the limit of large P\'{e}clet number (and large $q$), the tracer experiences linear advection that is independent of the diffusivity, as well as a non-local advection and non-local diffusion, both with a coefficient that is proportional to $\sqrt{\kappa U_0}$. Thus an increase to the background diffusivity associated with the across-jet mixing, increases the coefficient associated with the along-jet non-local diffusion. An attractive problem to consider is how flow instabilities or background turbulence associated with coherent jets can introduce spatial inhomogeneities to the background diffusivity, and whether a closed equation of the form (\ref{closed_eqn_large_Pe_q}) can be written for a new spatially varying $\kappa$ associated with the background turbulent flow. 
Lastly, it remains to be seen how the derived closures for large $\Pen$ values, which can be considered Eulerian, fair against diagnostic Lagrangian methods that provide instantaneous eddy diffusion coefficients even in the presence of chaotic advection \citep{nakamura1996two}.
\section{Conclusion}
\label{sec:conclusion}

In this study we have provided, to our knowledge, the first analytical solution to the problem of shear dispersion for all $t>0$. We thereby calculate exact closure operators (\ref{exact_closure}) for a wide range of averaging kernels, and derive analytical approximations to the exact closures (\ref{quadratic_operator}), which are expressed as the sum of two differential operators of fractional order. The approximations capture the behavior of the solution for large and small P\'{e}clet number, and are continuous in wavenumber space. These results establish the limits of validity of G. I. Taylor's effective diffusivity $\kappa^* \propto U_0^2/\kappa$, where $U_{0}$ is the characteristic flow speed and $\kappa$ is the molecular diffusivity. It is valid in the small P\'{e}clet number limit (strongly diffusive flows) for any initial tracer wavenumber $k>0$. It is also valid in the large P\'{e}clet number limit, provided that the wavenumber of the initial condition satisfies $2k\ll 1/\Pen$, where $\Pen$ is the P\'{e}clet number. This requirement guarantees a wide scale separation between the flow and the tracer field, and is consistent with homogenization theory. For wavenumbers that do not satisfy this requirement, we establish a new large P\'{e}clet number limit that is associated with propagating behavior in the kinematic solution. The closure is then composed of a linear advection operator plus a differential operator of fractional order (a semi-derivative) that compactly represents (non-local) diffusion and advection, with a coefficient proportional to $\sqrt{\kappa U_{0}}$. We find that closures in agreement with homogenization theory (at large P\'{e}clet number) cannot capture the early time evolution of the averaged tracer, but it is accurately described by the new large P\'{e}clet number regime.

\vspace{1cm}
\textbf{Acknowledgments}. The authors acknowledge helpful conversations with C. Meneveau and A. Mani at an early stage of the research. This project was funded by the Johns Hopkins University Institute for Data Intensive Engineering and Science Seed Funding Initiative and NSF grant 1835640. The code used to calculate Mathieu functions, eigenvalues, and solutions is available at \texttt{github.com/Mikejmnez/Adv\_Diff\_Supplemental\_Materials}.

\appendix
\section{Calculation of Mathieu eigenfunctions and characteristic values}\label{appA}
Following \citet{chaos2002mathieu, olver2010nist, ziener2012mathieu}, Mathieu functions are calculated from an infinite, tridiagonal eigenvalue system that arises after substituting the definition (\ref{ce2n}) into Mathieu's equation, and grouping by cosine terms. We get for each cosine term
\begin{align}\label{recurrence}
    \cos(0 y):  0 & = a_{2n}A_{0}^{(2n)} - q A_{2}^{(2n)} , \nonumber\\
    \cos(2y): 0 & = \left(-4 + a_{2n}\right)A_{2}^{(2n)} -q\left[2A_{0}^{(2n)}+A_{4}^{(2n)}\right] , \nonumber\\
    \cos(4y): 0 & =  \left(-16 +a_{2n}\right)A^{(2n)}_{4} - q\left[A_{2}^{(2n)} + A_{6}^{2n}\right]  ,\nonumber\\
    \cos(6y): 0 & =  \left(-36 +a_{2n}\right)A_{6}^{(2n)} - q\left[A_{4}^{(2n)} + A_{8}^{(2n)} \right] , \nonumber\\
    \cos(2ry): 0 & =  \left( -4 r ^2 +a_{2n}\right) A_{2r}^{(2n)} -q\left[A_{2r+2}^{(2n)} + A_{2r-2}^{(2n)} \right] .
\end{align}
Therefore, 
\begin{equation}\label{eig_system}
\setlength{\arraycolsep}{5pt}
\renewcommand{\arraystretch}{1.3}
\left[\begin{array}{ccccccc}
    0 & \sqrt{2}q & 0 & 0 & 0 & \cdots & \\
      \displaystyle
    \sqrt{2}q & 4 & q & 0 & 0 & & \\
      \displaystyle
    0 & q & 16 & q & 0 &  &\\
      \displaystyle
    0 & 0 & q & 36 & q &  & \\
      \displaystyle
    & & &\ddots & \ddots & \ddots&\\
      \displaystyle
    & & & & q & 4r^2 & q\\
      \displaystyle
    & & & & & & \ddots
    \end{array}\right] \left[\begin{array}{c}
    \sqrt{2}A_{0}^{(2n)}\\
    A_{2}^{(2n)}\\
    A_{4}^{(2n)}\\
    A_{6}^{(2n)}\\
    \vdots\\
    A_{2r}^{(2n)}\\
    \vdots
    \end{array}\right]= a_{2n}
\left[\begin{array}{c}
    \sqrt{2}A_{0}^{(2n)}\\
    A_{2}^{(2n)}\\
    A_{4}^{(2n)}\\
    A_{6}^{(2n)}\\
    \vdots\\
    A_{2r}^{(2n)}\\
    \vdots
    \end{array}\right] ,
\end{equation}
where $q=2ik\Pen$. The elements of the eigenvector are the Fourier coefficients that determine each Mathieu function $ \ce_{2n}(q, \tilde{y})$ as defined by (\ref{ce2n}). They satisfy the orthonormality relationship \citep{ziener2012mathieu}:
\begin{equation}\label{ortho}
    \sum_{n=0}^{\infty}A_{2p}^{(2n)}A_{2r}^{(2n)} = \delta_{pr} - \frac{\delta_{0p}\delta_{0r}}{2} .
\end{equation}
This formula is used to test convergence of the tridiagonal eigenvalue system (\ref{eig_system}), as it must be truncated to numerically calculate the eigenvalues and eigenfunctions. See (\ref{identity}) and (\ref{cosine_ce_identity}) for other Mathieu function identities.

\section{Asymptotic Approximations to Mathieu's Eigenvalues}\label{appB}
Expansions of the eigenvalues $a_{2n} (q)$ as a function of parameter $q=2ik\Pen$  give
\begin{align}\label{small_q_as}
    a_{0} &= 2(k\Pen)^2 + \frac{7}{8}(k\Pen)^4 - \frac{29}{36}(k\Pen)^6 +O((k\Pen)^{8}) ,\\
    a_{2} &= 4 - \frac{5}{4}(k\Pen)^2 - \frac{763}{864}(k\Pen)^4 + \frac{1002401}{1244160}(k\Pen)^6 + O((k\Pen)^{8}) , \\
     a_{4} &= 16 - \frac{2}{15}(k\Pen)^2 + \frac{433}{54000}(k\Pen)^4 - \frac{5701}{42525000}(k\Pen)^6 + O((k\Pen)^8) , \\
    a_{6} &= 36 -\frac{2}{35}(k\Pen)^2 + \frac{187}{2744000}(k\Pen)^4 + \frac{6743617}{1452124800000}(k\Pen)^6 +O((k\Pen)^8) ,
\end{align}
which are valid for $k\Pen{\ll1/2}$ (\citealt{olver2010nist} Ch. 28).
For small $q$, the limit $\lim_{q \rightarrow 0} a_{2n}(q) = (2n)^2$ ensures convergence of Mathieu functions into Fourier functions. For large $q$, the first four eigenvalues asymptote to
\begin{align}\label{large_q_as}
a_{0}, a_{2} &\sim 2\sqrt{k\Pen}-\dfrac{1}{4} \pm i\left(2\sqrt{k\Pen}-4k\Pen \right) , \\
   a_{4}, a_{6} &\sim 10\sqrt{k\Pen}-\dfrac{13}{4} \pm i\left(10\sqrt{k\Pen} - 4k\Pen \right)
   \label{large_q_as_46}
\end{align}
\citep{hunter1981eigenvalues,ziener2012mathieu}.

\bibliography{jfm-references}

\newcommand{\noop}[1]{}
\begin{thebibliography}{40}
\expandafter\ifx\csname natexlab\endcsname\relax\def\natexlab#1{#1}\fi
\def\au#1{#1} \def\ed#1{#1} \def\yr#1{#1}\def\at#1{#1}\def\jt#1{\textit{#1}}
  \def\bt#1{#1}\def\bvol#1{\textbf{#1}} \def\vol#1{#1} \def\pg#1{#1}
  \def\publ#1{#1}\def\arxiv#1{#1}\def\org#1{#1}\def\st#1{\textit{#1}}

\bibitem[Aris(1956)]{aris1956dispersion}
{\sc \au{Aris, Rutherford}} \yr{1956}  \at{On the dispersion of a solute in a
  fluid flowing through a tube}.  \jt{Proceedings of the Royal Society of
  London. Series A. Mathematical and Physical Sciences}  \bvol{235}~(1200),
  \pg{67--77}.

\bibitem[Arscott(2014)]{arscott2014periodic}
{\sc \au{Arscott, Felix~Medland}} \yr{2014} {\em Periodic differential
  equations: an introduction to {Mathieu}, {Lam{\'e}}, and allied functions\/},
  ,  \vol{vol.~66}.  \publ{Elsevier}.

\bibitem[Batchelor(1956)]{batchelor1956steady}
{\sc \au{Batchelor, Go~K}} \yr{1956}  \at{On steady laminar flow with closed
  streamlines at large reynolds number}.  \jt{Journal of Fluid Mechanics}
  \bvol{1}~(2),  \pg{177--190}.

\bibitem[Bender(1999)]{bender1999complex}
{\sc \au{Bender, Carl~M}} \yr{1999}  \at{The complex pendulum}.  \jt{Physics
  Reports}  \bvol{315}~(1-3),  \pg{27--40}.

\bibitem[Bender \& Boettcher(1998)]{bender1998real}
{\sc \au{Bender, Carl~M} \& \au{Boettcher, Stefan}} \yr{1998}  \at{Real spectra
  in non-{Hermitian Hamiltonians having PT} symmetry}.  \jt{Physical Review
  Letters}  \bvol{80}~(24),  \pg{5243}.

\bibitem[Blanch \& Clemm(1969)]{blanch1969double}
{\sc \au{Blanch, G} \& \au{Clemm, DS}} \yr{1969}  \at{The double points of
  {Mathieu’s} differential equation}.  \jt{Mathematics of Computation}
  \bvol{23}~(105),  \pg{97--108}.

\bibitem[Brimacombe {\em et~al.\/}(2020)Brimacombe, Corless \&
  Zamir]{brimacombe2020computation}
{\sc \au{Brimacombe, Chris}, \au{Corless, Robert~M} \& \au{Zamir, Mair}}
  \yr{2020}  \at{Computation and applications of {Mathieu} functions: A
  historical perspective}.  \jt{arXiv preprint arXiv:2008.01812} .

\bibitem[Camassa {\em et~al.\/}(2010)Camassa, Lin \&
  McLaughlin]{camassa2010exact}
{\sc \au{Camassa, Roberto}, \au{Lin, Zhi} \& \au{McLaughlin, Richard~M}}
  \yr{2010}  \at{The exact evolution of the scalar variance in pipe and channel
  flow}.  \jt{Communications in Mathematical Sciences}  \bvol{8}~(2),
  \pg{601--626}.

\bibitem[Chaos-Cador \& Ley-Koo(2002)]{chaos2002mathieu}
{\sc \au{Chaos-Cador, Lorea} \& \au{Ley-Koo, E}} \yr{2002}  \at{Mathieu
  functions revisited: matrix evaluation and generating functions}.
  \jt{Revista mexicana de f{\'\i}sica}  \bvol{48}~(1),  \pg{67--75}.

\bibitem[Chen(2006)]{chen2006speculative}
{\sc \au{Chen, Wen}} \yr{2006}  \at{A speculative study of 2/3-order fractional
  {Laplacian} modeling of turbulence: Some thoughts and conjectures}.
  \jt{Chaos: An Interdisciplinary Journal of Nonlinear Science}  \bvol{16}~(2),
   \pg{023126}.

\bibitem[Deconinck {\em et~al.\/}(2014)Deconinck, Trogdon \&
  Vasan]{deconinck2014method}
{\sc \au{Deconinck, Bernard}, \au{Trogdon, Thomas} \& \au{Vasan, Vishal}}
  \yr{2014}  \at{The method of {Fokas} for solving linear partial differential
  equations}.  \jt{siam REVIEW}  \bvol{56}~(1),  \pg{159--186}.

\bibitem[Gent \& Mcwilliams(1990)]{gent1990isopycnal}
{\sc \au{Gent, Peter~R} \& \au{Mcwilliams, James~C}} \yr{1990}  \at{Isopycnal
  mixing in ocean circulation models}.  \jt{Journal of Physical Oceanography}
  \bvol{20}~(1),  \pg{150--155}.

\bibitem[Gent {\em et~al.\/}(1995)Gent, Willebrand, McDougall \&
  McWilliams]{gent1995parameterizing}
{\sc \au{Gent, Peter~R}, \au{Willebrand, Jurgen}, \au{McDougall, Trevor~J} \&
  \au{McWilliams, James~C}} \yr{1995}  \at{Parameterizing eddy-induced tracer
  transports in ocean circulation models}.  \jt{Journal of Physical
  Oceanography}  \bvol{25}~(4),  \pg{463--474}.

\bibitem[Grooms {\em et~al.\/}(2012)Grooms, Smith \&
  Majda]{grooms2012multiscale}
{\sc \au{Grooms, Ian}, \au{Smith, K~Shafer} \& \au{Majda, Andrew~J}} \yr{2012}
  \at{Multiscale models for synoptic--mesoscale interactions in the ocean}.
  \jt{Dynamics of atmospheres and oceans}  \bvol{58},  \pg{95--107}.

\bibitem[Haynes \& Vanneste(2014)]{haynes2014dispersion}
{\sc \au{Haynes, PH} \& \au{Vanneste, J}} \yr{2014}  \at{Dispersion in the
  large-deviation regime. {Part I: Shear} flows and periodic flows}.  \jt{arXiv
  preprint arXiv:1401.6665} .

\bibitem[Heiss(2004)]{heiss2004exceptional}
{\sc \au{Heiss, WD}} \yr{2004}  \at{Exceptional points of non-{Hermitian}
  operators}.  \jt{Journal of Physics A: Mathematical and General}
  \bvol{37}~(6),  \pg{2455}.

\bibitem[Heiss(2012)]{heiss2012physics}
{\sc \au{Heiss, WD}} \yr{2012}  \at{The physics of exceptional points}.
  \jt{Journal of Physics A: Mathematical and Theoretical}  \bvol{45}~(44),
  \pg{444016}.

\bibitem[Hinch(1991)]{hinchperturbation}
{\sc \au{Hinch, EJ}} \yr{1991} {\em Perturbation Methods\/}.  \publ{Cambridge
  University Press and references therein}.

\bibitem[Holmes(2012)]{holmes2012introduction}
{\sc \au{Holmes, Mark~H}} \yr{2012} {\em Introduction to perturbation
  methods\/}, ,  \vol{vol.~20}.  \publ{Springer Science \& Business Media}.

\bibitem[Hunter \& Guerrieri(1981)]{hunter1981eigenvalues}
{\sc \au{Hunter, C} \& \au{Guerrieri, B}} \yr{1981}  \at{The eigenvalues of
  {Mathieu's} equation and their branch points}.  \jt{Studies in Applied
  Mathematics}  \bvol{64}~(2),  \pg{113--141}.

\bibitem[Lischke {\em et~al.\/}(2020)Lischke, Pang, Gulian, Song, Glusa, Zheng,
  Mao, Cai, Meerschaert, Ainsworth {\em et~al.\/}]{lischke2020fractional}
{\sc \au{Lischke, Anna}, \au{Pang, Guofei}, \au{Gulian, Mamikon}, \au{Song,
  Fangying}, \au{Glusa, Christian}, \au{Zheng, Xiaoning}, \au{Mao, Zhiping},
  \au{Cai, Wei}, \au{Meerschaert, Mark~M}, \au{Ainsworth, Mark} \& \au{others}}
  \yr{2020}  \at{What is the fractional {Laplacian}? a comparative review with
  new results}.  \jt{Journal of Computational Physics}  \bvol{404},
  \pg{109009}.

\bibitem[Majda \& Kramer(1999)]{majda1999simplified}
{\sc \au{Majda, Andrew~J} \& \au{Kramer, Peter~R}} \yr{1999}  \at{Simplified
  models for turbulent diffusion: theory, numerical modelling, and physical
  phenomena}.  \jt{Physics reports}  \bvol{314}~(4-5),  \pg{237--574}.

\bibitem[Mani \& Park(2019)]{mani2019macroscopic}
{\sc \au{Mani, Ali} \& \au{Park, Danah}} \yr{2019} Macroscopic forcing method:
  a tool for turbulence modeling and analysis of closures,  \arxiv{arXiv:
  1905.08342}.

\bibitem[Marshall \& Speer(2012)]{marshall2012closure}
{\sc \au{Marshall, John} \& \au{Speer, Kevin}} \yr{2012}  \at{Closure of the
  meridional overturning circulation through {Southern Ocean} upwelling}.
  \jt{Nature Geoscience}  \bvol{5}~(3),  \pg{171--180}.

\bibitem[McLachlan(1947)]{McLachlan1947Mathieu}
{\sc \au{McLachlan, N.W.}} \yr{1947} {\em Theory and Application of {Mathieu}
  Functions\/}.  \publ{Oxford University Press}.

\bibitem[Mercer \& Roberts(1990)]{mercer1990centre}
{\sc \au{Mercer, GN} \& \au{Roberts, AJ}} \yr{1990}  \at{A centre manifold
  description of contaminant dispersion in channels with varying flow
  properties}.  \jt{SIAM Journal on Applied Mathematics}  \bvol{50}~(6),
  \pg{1547--1565}.

\bibitem[Miri \& Alu(2019)]{miri2019exceptional}
{\sc \au{Miri, Mohammad-Ali} \& \au{Alu, Andrea}} \yr{2019}  \at{Exceptional
  points in optics and photonics}.  \jt{Science}  \bvol{363}~(6422).

\bibitem[Mulholland \& Goldstein(1929)]{mulholland1929}
{\sc \au{Mulholland, HP} \& \au{Goldstein, S}} \yr{1929}  \at{The
  characteristic numbers of the {Mathieu} equation with purely imaginary
  parameter}.  \jt{The London, Edinburgh, and Dublin Philosophical Magazine and
  Journal of Science}  \bvol{8}~(53),  \pg{834--840}.

\bibitem[Nakamura(1996)]{nakamura1996two}
{\sc \au{Nakamura, Noboru}} \yr{1996}  \at{Two-dimensional mixing, edge
  formation, and permeability diagnosed in an area coordinate}.  \jt{Journal of
  the atmospheric sciences}  \bvol{53}~(11),  \pg{1524--1537}.

\bibitem[Oldham \& Spanier(1974)]{oldham1974fractional}
{\sc \au{Oldham, Keith} \& \au{Spanier, Jerome}} \yr{1974} {\em The fractional
  calculus theory and applications of differentiation and integration to
  arbitrary order\/}.  \publ{Elsevier}.

\bibitem[Olver {\em et~al.\/}(2010)Olver, Lozier, Boisvert \&
  Clark]{olver2010nist}
{\sc \au{Olver, Frank W~J}, \au{Lozier, Daniel~W}, \au{Boisvert, Ronald~F} \&
  \au{Clark, Charles~W}} \yr{2010} {\em NIST handbook of mathematical functions
  hardback and {CD-ROM}\/}.  \publ{Cambridge University press}.

\bibitem[Park {\em et~al.\/}(2018)Park, Shirian \& Mani]{park2018macroscopic}
{\sc \au{Park, Danah}, \au{Shirian, Yasaman} \& \au{Mani, Ali}} \yr{2018}
  \at{Macroscopic forcing method and its application in assessment of {RANS}
  models}.  \jt{Bulletin of the American Physical Society}  \bvol{63}.

\bibitem[Podlubny(1998)]{podlubny1998fractional}
{\sc \au{Podlubny, Igor}} \yr{1998} {\em Fractional differential equations: an
  introduction to fractional derivatives, fractional differential equations, to
  methods of their solution and some of their applications\/}.
  \publ{Elsevier}.

\bibitem[Rhines \& Young(1983)]{rhines1983rapidly}
{\sc \au{Rhines, Peter~B} \& \au{Young, William~R}} \yr{1983}  \at{How rapidly
  is a passive scalar mixed within closed streamlines?}  \jt{Journal of Fluid
  Mechanics}  \bvol{133},  \pg{133--145}.

\bibitem[Samko {\em et~al.\/}(1993)Samko, Kilbas, Marichev {\em
  et~al.\/}]{samko1993fractional}
{\sc \au{Samko, Stefan~G}, \au{Kilbas, Anatoly~A}, \au{Marichev, Oleg~I} \&
  \au{others}} \yr{1993} {\em Fractional integrals and derivatives\/}, ,
  \vol{vol.~1}.  \publ{Gordon and Breach Science Publishers, Yverdon
  Yverdon-les-Bains, Switzerland}.

\bibitem[Smith(2005)]{smith2005tracer}
{\sc \au{Smith, K~Shafer}} \yr{2005}  \at{Tracer transport along and across
  coherent jets in two-dimensional turbulent flow}.  \jt{Journal of Fluid
  Mechanics}  \bvol{544},  \pg{133--142}.

\bibitem[Taylor(1953)]{taylor1953dispersion}
{\sc \au{Taylor, Geoffrey~Ingram}} \yr{1953}  \at{Dispersion of soluble matter
  in solvent flowing slowly through a tube}.  \jt{Proceedings of the Royal
  Society of London. Series A. Mathematical and Physical Sciences}
  \bvol{219}~(1137),  \pg{186--203}.

\bibitem[Tseng {\em et~al.\/}(2000)Tseng, Pei \& Hsia]{tseng2000computation}
{\sc \au{Tseng, Chien-Cheng}, \au{Pei, Soo-Chang} \& \au{Hsia, Shih-Chang}}
  \yr{2000}  \at{Computation of fractional derivatives using fourier transform
  and digital fir differentiator}.  \jt{Signal Processing}  \bvol{80}~(1),
  \pg{151--159}.

\bibitem[Young \& Jones(1991)]{young1991shear}
{\sc \au{Young, WR~a} \& \au{Jones, Scott}} \yr{1991}  \at{Shear dispersion}.
  \jt{Physics of Fluids A: Fluid Dynamics}  \bvol{3}~(5),  \pg{1087--1101}.

\bibitem[Ziener {\em et~al.\/}(2012)Ziener, R{\"u}ckl, Kampf, Bauer \&
  Schlemmer]{ziener2012mathieu}
{\sc \au{Ziener, CH}, \au{R{\"u}ckl, M}, \au{Kampf, T}, \au{Bauer, WR} \&
  \au{Schlemmer, Heinz-Peter}} \yr{2012}  \at{Mathieu functions for purely
  imaginary parameters}.  \jt{Journal of Computational and Applied Mathematics}
   \bvol{236}~(17),  \pg{4513--4524}.

\end{thebibliography}
\bibliographystyle{jfm}

\end{document}